\documentclass[english,a4paper,twoside,twocolumn,superscriptaddress,groupedaddress,showpacs]{revtex4}
\usepackage[T1]{fontenc}
\usepackage[latin9]{inputenc}
\usepackage{graphicx}
\usepackage{amssymb}

\makeatletter

\providecommand{\tabularnewline}{\\}
\newcommand{\lyxdot}{.}

\makeatletter

\@ifundefined{definecolor}
 {\@ifundefined{definecolor}
 {\usepackage{color}}{}
}{}

\makeatletter

\@ifundefined{definecolor}
 {\@ifundefined{definecolor}
 {\@ifundefined{definecolor}
 {\usepackage{color}}{}
}{}
}{}

\makeatletter



\makeatother

\makeatother

\makeatother

\makeatother

\makeatother

\usepackage{babel}

\begin{document}

\title{Phase separation and pairing regimes in the one-dimensional asymmetric
Hubbard model}

\author{L. Barbiero}

\affiliation{Dipartimento di Fisica del Politecnico, corso Duca degli Abruzzi,
24, 10129, Torino, Italy}

\author{M. Casadei}

\affiliation{Fritz Haber Institute of the Max Planck Society, Abteilung Theorie,
Faradayweg 4-6, 14195, Berlin, Germany}

\author{M. Dalmonte}

\affiliation{Dipartimento di Fisica dell'Università di Bologna, via Irnerio, 46,
40126, Bologna, Italy}

\affiliation{Sezione INFN di Bologna}

\author{C. Degli Esposti Boschi}

\affiliation{CNR, Unità CNISM di Bologna}

\affiliation{Dipartimento di Fisica dell'Università di Bologna, viale Berti-Pichat,
6/2, 40127, Bologna, Italy}

\author{E. Ercolessi}

\affiliation{Dipartimento di Fisica dell'Università di Bologna, via Irnerio, 46,
40126, Bologna, Italy}

\affiliation{Sezione INFN di Bologna}

\author{F. Ortolani}

\affiliation{Dipartimento di Fisica dell'Università di Bologna, via Irnerio, 46,
40126, Bologna, Italy}

\affiliation{Sezione INFN di Bologna}

\begin{abstract}
We address some open questions regarding the phase diagram of the
one-dimensional Hubbard model with asymmetric hopping coefficients
and balanced species. In the attractive regime we present a numerical
study of the passage from on-site pairing dominant correlations at
small asymmetries to charge-density waves in the region with markedly
different hopping coefficients. In the repulsive regime we exploit
two analytical treatments in the strong- and weak-coupling regimes
in order to locate the onset of phase separation at small and large
asymmetries respectively. 
\end{abstract}


\maketitle

\section{Introduction}

In this paper we study a variation of the one-dimensional Hubbard
model (HM) in which the difference between the hopping amplitudes,
say $t_{\uparrow}>t_{\downarrow}$, is responsible for an explicit
breaking of the rotational symmetry. It is described by the Hamiltonian

\begin{equation}
H=-\sum_{j\sigma}t_{\sigma}(c_{j\sigma}^{\dagger}c_{j+1\sigma}+{\rm h.c.})+U\sum_{j}n_{j\uparrow}n_{j\downarrow}\label{eq:AHM}\end{equation}
 where $c_{j\sigma}$ denotes the annihilation operator of a fermion
with $\sigma=\uparrow,\downarrow$ at site $j$ and $n_{j\sigma}=c_{j\sigma}^{\dagger}c_{j\sigma}$
are the associated number operators.

This asymmetric Hubbard model (AHM) has been studied in the past \cite{FK1969}
to describe the essential features of the metal-insulator transition
in rare-earth materials and transition-metal oxides; in this case
$\sigma$ represents two types of spinless fermions (the real spin
being considered not essential for the transition to be modelled).
The {}``light'' particles are described by band (Bloch) states,
while the {}``heavy'' ones tend to be localized on lattice (Wannier)
sites. More recently, this model has gained a renewed interest in
experiments with optical lattices, in which both the effective strengths
of the kinetic and of the potential parts can be varied in a rather
controlled way, including the possibility of reaching the attractive
regime $U<0$. The possibility to use cold atoms \cite{BDZ2008,Jetal1998}
to engineer condensed matter systems with a high tunability offers
an experimental way to test theoretical results with great accuracy.
Two-species models with different hopping coefficients can also be
realized by trapping atomic clouds with two internal states of different
angular momentum, thereby introducing a spin dependent optical lattice,
which enables to modify the ratio $a=t_{\downarrow}/t_{\uparrow}$
by controlling the depth of the optical lattice \cite{LWZ2004}.
Yet another possibility is to trap two different species of fermionic
atoms, so that the {}``anisotropy'' $a$ is given naturally by the
ratio of masses. In the context of cold atoms in optical lattices
the subscripts $\sigma=\uparrow,\downarrow$ are not related to the
electron spin but label the two different species of fermions, either
different atoms with half-integer spin or different excited states
of one atomic specie with fine structure splitting.

\begin{figure}
\includegraphics[width=7cm]{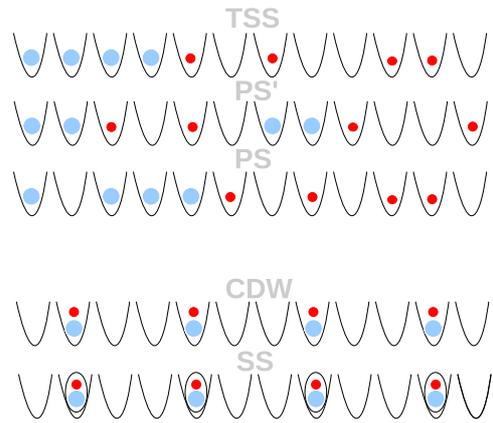}

\caption{Schematic representation of the phases discussed in the text. Large
and small disks represent heavy and light fermions respectively. The
ellipsis encircling the pairs in the SS phase denote onsite pairing.
The HP state defined in Eq. (\ref{eq:PsiE}) has not been drawn in
real space because it is defined directly in momentum space by filling
the two noninteracting bands up to $k_{F}$.\label{fig:cartoon} }

\end{figure}

Many recent papers on the subject are devoted to the onset of the
so-called Fulde-Ferrell-Larkin-Ovchinnikov (FFLO) phase, which may
occur with \textit{unbalanced} species \cite{BWHR2009,WCDS2009}.
In this paper we will consider instead the case of \textit{balanced}
species: $N_{\uparrow}=N_{\downarrow}$. The parameters that influence
the phase diagram can be cast in the form of an anisotropy coefficient
$z=(t_{\uparrow}-t_{\downarrow})/(t_{\uparrow}+t_{\downarrow})=(1-a)/(1+a)$
and a dimensionless onsite potential $u=U/t$, where $t=(t_{\uparrow}+t_{\downarrow})/2$
is an overall energy scale. In addition one can consider the effect
of the total filling $n=N/L$, $N$ being the total number of fermions
and $L$ the chain length. At variance with the typical situation
in condensed matter physics, where the bulk filling and magnetization
$m^{z}=(n_{\uparrow}-n_{\downarrow})/2$ are controlled in a grand-canonical
framework by the chemical potential and an external magnetic field
respectively, in the context of cold atoms it is conceivable to fix
independently the number of particles in each of the two species,
despite the fact that only a given choice of $ $the densities $n_{\uparrow}$
and $n_{\downarrow}$ might correspond to the absolute minimum of
the grand potential. Henceforth we will assume that they are equally
populated, that is $n_{\uparrow}=n_{\downarrow}=n/2$ and we will
limit ourselves to $n<1$. Note that if $E(n,m^{z})$ is the energy
in a given sector of $n$ and $m^{z}$ (these two quantities are good
quantum numbers also in the asymmetric case), then a particle-hole
transformation $n_{j\sigma}\to(1-n_{j\sigma})$ leads to $E(2-n,-m^{z})=E(n,m^{z})-UL(n-1)$,
so that for $m^{z}=0$ it is sufficient to analyze the phase diagram
for positive and negative $U$ and $n<1$, in order to infer features
also at $n>1$. The phase diagram at half-filling, $n=1$, has been
studied in ref. \cite{FDL1995}, soon after the development of White's
density-matrix renormalization group (DMRG) method \cite{W1993}.
The limiting case of the symmetric HM $(z=0)$ can be solved exactly
via the Bethe \textit{ansatz} approach \cite{LW1968}. The opposite
extremal case $z=1$ is usually called the Falicov-Kimball (FK) model.
In any dimensionality, it has been proved \cite{FLU2002} that for
large \textit{positive} $U$ and $n\neq$1 the system has a ground
state characterized by a spatially non-homogeneous density profile
where the heavy particles are compressed in a region with length $L_{\downarrow}=N_{\downarrow}$.
In this paper we refer to this phase as \textit{totally segregated
state} (TSS). A similar phase, where the heavy particles acquire a
finite kinetic energy ($L_{\downarrow}>N_{\downarrow}$), appears
for $z\lesssim1$ \cite{U2004} and can be interpreted as the result
of an effective \textit{attractive} interaction between light fermions,
mediated by the heavy ones (see \cite{DLE1995} in 1D and \cite{DLF1996}
in 2D). The latter situation is denoted in the following as \textit{phase
separation} (PS). A cartoon representation of the different phases
is presented in Fig. \ref{fig:cartoon}. At smaller asymmetries, the
system is instead in a more conventional spatially homogeneous phase
(HP). The ground state phase diagram for the model in the $m^{z}=0$
sector has been discussed in \cite{CHG2005} by means of the bosonization
approach. In this context \cite{WCG2007} the HP-PS transition line
at $u>0$ has been interpreted as the curve in $(z,u)$-plane where
the velocity of one of the two decoupled bosonic modes vanishes.

Let us summarize the content of this paper. In Sec. \ref{sec:uneg}
we will study numerically the attractive regime ($U<0$), by using
a DMRG program. In particular we will examine which kind of correlations
(charge or pairing) is dominant in this region of the phase diagram.
Then we will move to consider the repulsive ($U>0$) regime. In Sec.
\ref{sec:w-c} we will analytically discuss the weak coupling regime
($|U|\ll t_{\sigma}$) by means of a variational method that compares
the energy of the PS state with that of the HP state, the latter being
calculated within a second order perturbation analysis. In Sec. \ref{sec:s-c}
we move to study the strong coupling regime ($U\gg t_{\sigma}$) in
order to determine a phase diagram which includes also different types
of PS states. The results are summarized and the conclusions are drawn
in Sec. \ref{sec:conc}. %

\section{Singlet-pairing to charge-density wave transition at $U<0$\label{sec:uneg}}

\begin{figure}
\includegraphics[clip,height=7cm]{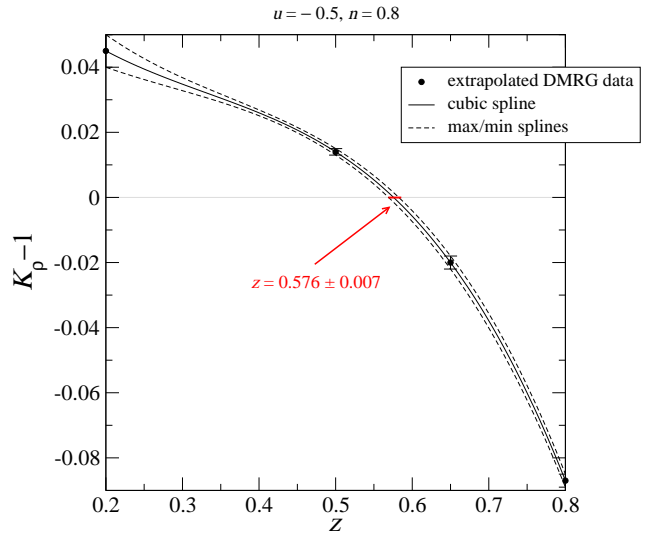}

\caption{Example of cubic splines interpolation to locate the transition point
and the associated error, using the DMRG data of table \ref{tab:KrhoUneg}.
The horizontal solid segment at $K_{\rho}=1$ indicates the reported
interval.\label{fig:splines}}

\end{figure}

\begin{figure}
\includegraphics[clip,height=7cm]{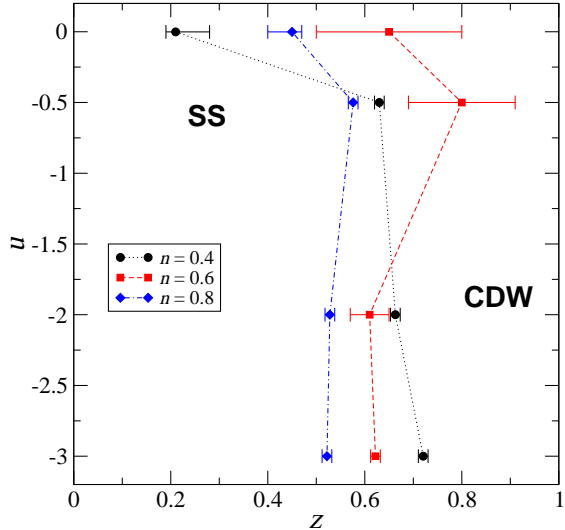}

\caption{Numerical estimates of the transition lines from SS (on the left)
to CDW (on the right) dominant correlations in the attractive regime
for the values of filling reported in the legend. Not all the data
used to obtain the figure have been reported in table \ref{tab:KrhoUneg}.
The error bars associated to the transition points have been determined
by means of splines passing through the upper and lower edges of each
interval of possible values of $K_{\rho}$ reported in the table (see
the construction of Fig. \ref{fig:splines}); when the uncertainty
in $z$ turned to be smaller than $10^{-2}$ we have conventionally
set it to this value to account for the approximation introduced by
cubic splines interpolation. The lines joining the points are guides
for the eye.$ $\label{fig:sscdw}}

\end{figure}

One of the open points raised in ref. \cite{CHG2005} is the existence
of regions at $U<0$ characterized by dominating charge-density wave
(CDW) correlations instead of the singlet-superconducting (SS) ones
that one has in the attractive symmetric HM. On a lattice the CDW
and the SS correlation functions are defined respectively as \[
C(r)=\langle n_{j}n_{j+r}\rangle-\langle n_{j}\rangle\langle n_{j+r}\rangle\]
 \[
P(r)=\langle\eta_{j}^{\dagger}\eta_{j+r}\rangle\]
 where $n_{j}=n_{j\uparrow}+n_{j\downarrow}$ while $\eta_{j}=c_{j\downarrow}c_{j\uparrow}$
is the operator that destroys an onsite pair with singlet spin wavefunction.
At large distances $r$, bosonization procedures predict \cite{Gia2003}\[
C(r)\sim\frac{-K_{\rho}}{\pi^{2}r^{2}}+A\frac{\cos(2k_{F}r)}{r^{K_{\rho}+K_{\sigma}}}\]
 \[
P(r)\sim\frac{B}{r^{1/K_{\rho}+K_{\sigma}}}\]
 where $A$ and $B$ are constants, and $K_{\rho}$ and $K_{\sigma}$
are the Luttinger parameters for the charge and spin degrees of freedom
respectively. Clearly $C(r)$ dominates over $P(r)$ when $K_{\rho}<1$,
while $K_{\sigma}$ has to be fixed to 0 for gapped spin phases. A
numerical estimate of $K_{\rho}$ from finite-size data can be obtained
as in ref. \cite{SBC2004} by considering the structure factor\[
S(q)=\sum_{r}{\rm e}^{{\rm i}qr}\langle n_{j}n_{j+r}\rangle\]
 Here we have dropped the dependence on $j$ because we implicitly
assume that the correlation functions are translationally invariant
due to periodic boundary conditions (PBC). The value $S(q=0)$ corresponds
to the average correlation and it diverges in the thermodynamic limit,
since typically $\langle n_{j}n_{j+r}\rangle$ saturates at a finite
value $n^{2}$ for large distances. So one may consider the connected
charge correlation $C(r)$ in order to avoid this divergence. This
choice affects the structure factor only at $q=0$ and bosonization
predicts that the value of $K_{\rho}$ is directly related to the
\textit{limit} $q\to0$\[
S(q)\simeq K_{\rho}\frac{q}{\pi}+\dots\]
 Following the procedure of ref. \cite{SBC2004} by selecting the
smallest possible non-vanishing momentum compatible with PBC $q_{1}=2\pi/L$
one builds a sequence that approximates the linear slope \[
K_{\rho}(L)=S(q_{1})\frac{L}{2}.\]
 The dependence on $L$ indicates that the sequence has to be extrapolated
to $L\to\infty$ in order to obtain the limit $q\to0$ and the parameter
$K_{\rho}$. The DMRG results corresponding to various fillings $n=0.4,\;0.6,\;0.8$
with $z=0.2,\;0.5,\;0.65,\;0.8$ and $u=-0.001,\;-0.5,\;-3$ are reported
in table \ref{tab:KrhoUneg}; in the caption we have reported also
the relevant features of our DMRG numerical calculations. For a fixed
value of $u<0$ we always find that the extrapolated $K_{\rho}$ decreases
with increasing $z$. From this grid of points we can have an idea
of the SS-CDW transition curve by locating the points at which $K_{\rho}=1$.
This has been done interpolating the data with splines. An example
of this procedure is given in Fig. \ref{fig:splines}, while the global
results are plotted in Fig. \ref{fig:sscdw}.

In ref. \cite{CHG2005} (Fig. 1 therein) the authors report a triangular
region obtained by means of bosonization at $u<0$ where the dominant
correlations are CDW or SS depending on the filling. Here we observe
that the shape of the transition line is indeed dependent on $n$:
for $u<-0.5$ the separation line might have both a negative and a
positive slope, depending on the value of $n$. For small (negative)
$u$ the curve has always a negative slope. Because of the uncertainty
related to DMRG and finite-size effects, our estimate of the transition
points is limited to $u\le-10^{-3}$; moving closer to $u=0$ would
produce values of $K_{\rho}$ essentially always equal to 1 within
the numerical error, for all values of $z$, so we have not pushed
our analysis and conclusions closer to $u=0$.

Finally we should mention that a direct inspection of the charge correlation
functions in real and in Fourier space reveals that the only characteristic
wavenumber is $2k_{F}=\pi n$, where typically the structure factor
displays a peak for $u<0$, but there is neither FFLO behavior - as
expected since we have selected balanced species - nor a collapse
(predicted at sufficiently large negative $u$ \cite{M2007}).

\begin{table*}
\begin{tabular}{|c|c|c|c|c|c|}
\hline 
\textbf{$u$ \textbackslash{} $z$}  &  & \textbf{$0.2$}  & \textbf{$0.5$}  & \textbf{$0.65$}  & \textbf{$0.8$}\tabularnewline
\hline
\hline 
 & \textbf{$n=0.4$}  &  &  &  & \tabularnewline
\hline 
\textbf{$-10^{-3}$}  &  & \textbf{$1.001\pm0.002^{\ddagger}$}  & \textbf{$0.9978\pm0.0002^{*}$}  & \textbf{$0.994\pm0.001^{*}$}  & \textbf{$0.986\pm0.004^{*}$}\tabularnewline
\hline 
\textbf{$-0.5$}  &  & \textbf{$1.0700\pm0.0004$}  & \textbf{$1.038\pm0.001$}  & \textbf{$0.992\pm0.002$}  & \textbf{$0.906\pm0.003$}\tabularnewline
\hline 
\textbf{$-3$}  &  & \textbf{$1.4209\pm0.0005$}  & \textbf{$1.2528\pm0.0003$}  & \textbf{$1.0976\pm0.0002$}  & \textbf{$0.86715\pm0.00005$}\tabularnewline
\hline
\hline 
 & \textbf{$n=0.6$}  &  &  &  & \tabularnewline
\hline 
\textbf{$-10^{-3}$}  &  & \textbf{$1.0034\pm0.0008^{*}$}  & \textbf{$1.001\pm0.001^{\ddagger}$}  & \textbf{$1.000\pm0.002^{\ddagger}$}  & \textbf{$0.998\pm0.004^{\ddagger}$}\tabularnewline
\hline 
\textbf{$-0.5$}  &  & \textbf{$1.050\pm0.001^{\ddagger}$}  & \textbf{$1.04\pm0.01^{\ddagger}$}  & \textbf{$1.03\pm0.02^{\ddagger}$}  & \textbf{$1.00\pm0.05^{\ddagger}$}\tabularnewline
\hline 
\textbf{$-3$}  &  & \textbf{$1.26016\pm0.00007$}  & \textbf{$1.110\pm0.003$}  & \textbf{$0.969\pm0.004$}  & \textbf{$0.751\pm0.004$}\tabularnewline
\hline
\hline 
 & \textbf{$n=0.8$}  &  &  &  & \tabularnewline
\hline 
\textbf{$-10^{-3}$}  &  & \textbf{$1.008\pm0.002^{*}$}  & \textbf{$0.9976\pm0.0001^{*}$}  & \textbf{$0.983\pm0.005$}  & \textbf{$0.951\pm0.009$}\tabularnewline
\hline 
\textbf{$-0.5$}  &  & \textbf{$1.045\pm0.005^{\ddagger}$}  & \textbf{$1.014\pm0.001$}  & \textbf{$0.980\pm0.002$}  & \textbf{$0.913\pm0.002$}\tabularnewline
\hline 
\textbf{$-3$}  &  & \textbf{$1.1603\pm0.0001$}  & \textbf{$1.0163\pm0.0008$}  & \textbf{$0.880\pm0.001$}  & \textbf{$0.6722\pm0.0008$}\tabularnewline
\hline
\end{tabular}

\caption{Extrapolations for the parameter $K_{\rho}$ from DMRG simulations
with PBC, $L=10,20,30,40,50$, 1100-1300 optimized states and seven
finite-system sweeps. This conservative choice guarantees an energy
relative error of $O(10^{-6})$ up to $L=30$ and $O(10^{-5})$ up
to $L=50$ (recall that the charge degrees of freedom are always gapless).
Unless otherwise specified the extrapolations have been performed
using quadratic fits in $1/L$ and the error bars are evaluated according
to \cite{NR} (Chapter 15) using the sum of squared differences normalized
to the fit degrees of freedom as a measure of the spread in the ordinates.
$\ddagger$: Oscillating about reported value with spread. {*}: Linear
fit in $1/L$.\label{tab:KrhoUneg}}

\end{table*}

\section{Weak-coupling limit\label{sec:w-c}}

From a quantitative point of view, bosonization cannot be conclusive
about the location of the HP to PS transition at $z\cong1$. This
is due to the fact that, while a continuum limit approach is justified
close to $z=0$, in the FK limit even a small $U$ can involve processes
away from the Fermi surface, and so the requirement $|U|\ll\min[t_{\uparrow},t_{\downarrow}]$
strictly reduces the range of reliability of this approach\cite{CHG2005}.

Numerical data \cite{F2007} indicate that in the highly asymmetric
regime phase separation appears above a critical value of $U$ which
approaches zero in the low-density limit.

When approaching the FK limit $z\rightarrow1$, the kinetic energy
of the lighter species becomes the dominant term of the Hamiltonian
at weak coupling, $U\ll t_{\uparrow}$. For $U>0$, the competition
between the light particle kinetic energy and the repulsive on-site
interaction can drive the system into an instability with respect
to phase separation \cite{U2004}: there exists a critical value
$u_{PS}(z,n)$ above which light particles will occupy a large region
of the system where $n_{i\downarrow}=0$, creating an effective pressure
which confines heavy particles into a small region with density close
to 1; this effect is reminiscent of the segregated regime present
in the FK model in the repulsive regime.

Different numerical and analytical methods have been proposed in literature
to identify this phase transition between the HP and the PS/TSS regimes
\cite{WCG2007,Cetal2008,F2007}. In the following, we will (\textit{i})
apply a second order perturbation theory approach, first introduced
\cite{MV1989} to study the ground state of the weakly interacting
symmetric HM, to compute the energy of a homogeneous phase ground
state of the AHM and (\textit{ii}) compare HP and PS or TSS energies
in order to detect the line of quantum phase transition as a function
of the original model parameters $z,u,n$. Finally, a comparison with
previous results will be presented.

\subsection{Trial wave functions\label{sub:twfw-c}}

In the low coupling regime, we can consider as extended ground state
the exact one at $U=0$ (and $t_{\downarrow}\neq0)$, which can be
obtained by filling both Fermi bands up to $k_{F\sigma}$: \begin{equation}
|\Psi\rangle_{HP}=\prod_{|q|\le k_{F\downarrow}}\tilde{c}_{q\downarrow}^{\dagger}\prod_{|k|\le k_{F\uparrow}}\tilde{c}_{k\uparrow}^{\dagger}|0\rangle\label{eq:PsiE}\end{equation}
 where $\vert0\rangle$ represents the zero-fermions vacuum and $\tilde{c}_{k\sigma}^{\dagger}$
are the creation fermionic operators in Fourier space. While $|\psi\rangle_{HP}$
is not an exact eigenstate of the full Hamiltonian, perturbation theory
above this ansatz have provided excellent results for $z=0$ \cite{MV1989},
where a comparison with the exact solution is possible, and, as shown
later in the section, even in the highly asymmetric regime.

The PS ground state instead can be obtained in the following way:
first, we confine all heavy particles in a given part of the lattice
of relative length $\nu$, with $(L-N_{\uparrow})/L\ge\nu\ge N_{\downarrow}/L$;
then we consider two different chains of length $L_{\downarrow}=\nu L,\; L_{\uparrow}=L(1-\nu)$
respectively and then fill the new Fermi bands till the momenta $k_{F\sigma}'=\pi N_{\sigma}/L_{\sigma}$:

\[
|\Psi(\nu)\rangle_{PS}=\prod_{|q'|\le k_{F\downarrow}'}\tilde{c}_{q'\downarrow}^{\dagger}\prod_{|k'|\le k_{F\uparrow}'}\tilde{c}_{k'\uparrow}^{\dagger}|0\rangle.\]
 In practice we have to consider that the effective light-particle
and heavy-particle {}``chain lengths'' are not $L$ but $L(1-\nu)$
and $L\nu$, respectively.

In addition, we can define a TSS as the one with a completely full
region of heavy particles, i.e. $\nu=N_{\downarrow}/L$. In this case,
the variational wave function can be written as: \[
|\Psi\rangle_{TSS}=\prod_{(1-n_{\downarrow})L<j<L}c_{j\downarrow}^{\dagger}\prod_{|k'|\le k_{F\uparrow}'}\tilde{c}_{k'\uparrow}^{\dagger}|0\rangle.\]
 where now the down-fermion creation operators are taken in real space
representation. Both TSS and PS state trial wave functions are eigenstate
of the Hamiltonian up to a boundary term which we neglect in the following
$L\rightarrow\infty$ limit.

\subsection{Ground state energies\label{sub:ground stateew-c}}

The instability of a homogeneous ground state towards a phase separated
one can be analyzed by computing the corresponding zero temperature
energy: \[
\mathcal{E}_{PS}=\frac{_{PS}\langle\Psi\vert H|\Psi\rangle_{PS}}{L},\quad\mathcal{E}_{HP}=\frac{_{HP}\langle\Psi\vert H|\Psi\rangle_{HP}}{L}\]
 and by comparing them to get the phase transition hypersurface in
parameter space described by:\[
\mathcal{E}_{PS}(n_{\uparrow},n_{\downarrow},U,z)-\mathcal{E}_{HP}(n_{\uparrow},n_{\downarrow},U,z)=0\]
A similar criterion can be applied to distinguish between TSS and
PS state without segregation, as described later in the section. We
remark that in the PS region, due to the fact that within the two
subchains of length $L_{\uparrow,\downarrow}$ the up and down particles
do not overlap, the interaction term provides at most a boundary contribution
which can be neglected in the thermodynamic limit. We will come back
to this point later.

We can then compute the PS state energy density $\mathcal{E}_{PS}$
considering only the kinetic term contribution. For a general $\nu$,
the result is \begin{equation}
\mathcal{E}_{PS}(\nu)=-t(1+z)\frac{2(1-\nu)}{\pi}\sin\left(\frac{\pi n_{\uparrow}}{1-\nu}\right)\label{eq:eps}\end{equation}
 \[
-t(1-z)\frac{2\nu}{\pi}\sin\left(\frac{\pi n_{\downarrow}}{\nu}\right)\]
As already stated in \cite{U2004} (in particular Sec. 3 therein),
it is possible to fix the lowest energy state with respect to $\nu$
searching for a minimum of (\ref{eq:eps}) at fixed densities and
hopping rates. The corresponding condition $\partial_{\nu}\mathcal{E}_{PS}=0$
becomes\[
(1+z)\sin\left(\frac{\pi n_{\uparrow}}{1-\nu}\right)-\frac{\pi n_{\uparrow}}{1-\nu}\cos\left(\frac{\pi n_{\uparrow}}{1-\nu}\right)\]
 \begin{equation}
=(1-z)\left[\sin\left(\frac{\pi n_{\downarrow}}{\nu}\right)-\frac{\pi n_{\downarrow}}{\nu}\cos\left(\frac{\pi n_{\downarrow}}{\nu}\right)\right]\label{eq:znuunb}\end{equation}
If there exists a value of $\nu$, $\nu^{*}$ with $1-n_{\uparrow}>\nu^{*}>n_{\downarrow}$,
which satisfies this condition, then the lowest energy state is $|\Psi(\nu^{*})\rangle_{PS}$;
otherwise, the minimum of $\mathcal{E}_{PS}$ lies on the boundary
$\nu^{*}=n_{\downarrow}$ and TSS is energetically more favorable.
The boundary between these two regions is described by the condition
$\partial{\cal E}_{PS}/\partial\nu\vert_{\nu=n_{\downarrow}}=0$;
the solution of this equation provides a characteristic anisotropy
coefficient $\bar{z}$ for given $n_{\uparrow,\downarrow}$ that turns
out to be independent of $u$. We expect that $\bar{z}$ represents
a good estimate for the phase transition between TSS and $\nu\neq n_{\downarrow}$
states even at intermediate couplings. In fact, we find good agreement
with the values corresponding to the (almost) horizontal lines plotted
in Fig. 3 of ref. \cite{F2007}.

\begin{figure}
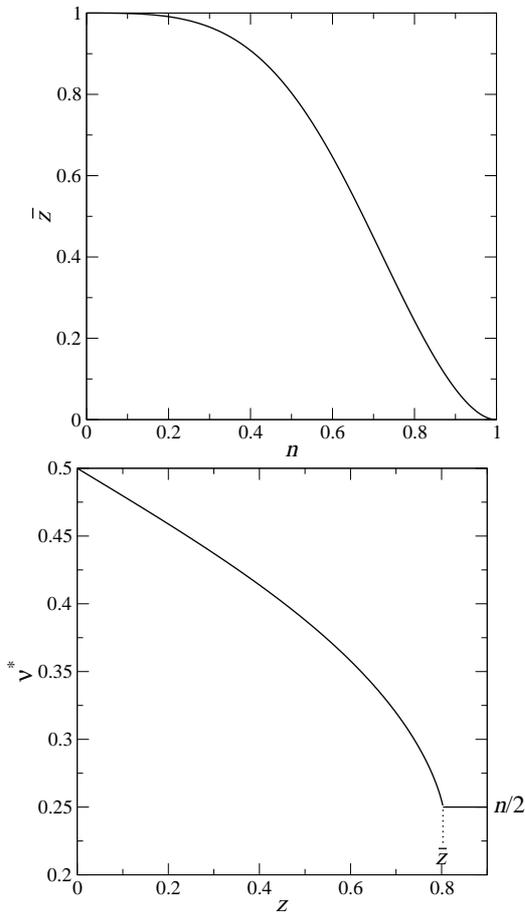

\includegraphics[clip,height=6cm]{zb_vs_n}

\includegraphics[clip,height=6cm]{nus_n05}

\caption{Upper panel: Plot of the function $\bar{z}(n)$. Lower panel: Optimal
value of $\nu$ as a function of $z$ in the whole interval $[0,1]$
for filling $n=1/2$.$ $\label{fig:zbnus}}

\end{figure}

As said before, here we consider only the balanced case $n_{\uparrow}=n_{\downarrow}=n/2$,
for which the condition that yields the optimal $\nu=\nu^{*}(z)$
in the range $\nu\in(n/2,1-n/2)$ simplifies to\begin{equation}
z=\frac{f\left(\frac{\pi n}{2\nu}\right)-f\left(\frac{\pi n}{2(1-\nu)}\right)}{f\left(\frac{\pi n}{2\nu}\right)+f\left(\frac{\pi n}{2(1-\nu)}\right)}\label{eq:znu}\end{equation}
with $f(x)=\sin x-x\cos x$. The upper limit for $\nu$ is formally
obtained by compressing the up-fermions so that $1-\nu=n/2$. When
$\nu=1/2$ the numerator of the right-hand side vanishes and $z$
reaches the minimum possible value $z=0$. When $\nu=\nu_{min}=n/2$
the right-hand side has the value\[
\bar{z}(n)=\frac{\pi-f(\pi n/(2-n))}{\pi+f(\pi n/(2-n))}.\]
This function decreases monotonically with the filling from $\bar{z}(0)=1$
to $\bar{z}(1)=0$. If $z\ge\bar{z}(n)$, the minimum energy is attained
at the lower limit $\nu^{*}=n/2$, corresponding to maximum compression
of the down-fermions, independently of the value of $z$. Clearly,
high densities favor a TSS state, since a large amount of light particles
produce a sufficient pressure in order to compress all heavy particles
in a very small region. When the density is low enough, this condition
cannot be satisfied, except at very large mass imbalances, and heavy
particles still contribute to the kinetic energy of the system. Fig.
\ref{fig:zbnus} shows an example of such a construction for $n=1/2$.

\begin{figure}
\includegraphics[height=4cm]{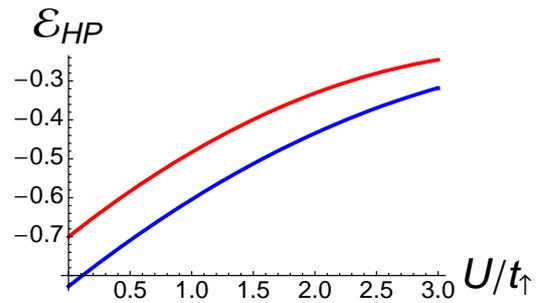}

\caption{{\small Energy density of the HP state at $n=1$ for different asymmetry
parameters: upper red line $t_{\downarrow}/t_{\uparrow}=0.1$, lower
blue line $t_{\downarrow}/t_{\uparrow}=0.3$.}$ $\label{fig:edensity}}

\end{figure}

The energy density of the HP state receives instead both kinetic and
interaction contributions: $\mathcal{E}_{HP}=\mathcal{E}_{T}+\mathcal{E}_{U}$.
The kinetic part is equivalent to the ground state energy density
of the non-interacting case\begin{equation}
\mathcal{E}_{T}=-t(1+z)\frac{2}{\pi}\sin(\pi n_{\uparrow})-t(1-z)\frac{2}{\pi}\sin(\pi n_{\downarrow})\label{eq:ET}\end{equation}
 whereas the interacting part in the weak coupling limit can be expressed
as a series in $U$ applying a second order perturbation theory \cite{MV1989}:
\[
\mathcal{E}_{U}=\frac{1}{L}\,_{HP}\langle\Psi|H_{U}|\Psi\rangle_{HP}\]
 \[
+\frac{1}{L}\left(_{HP}\langle\Psi|H_{U}\frac{1}{\mathcal{E}_{T}-H_{0}}H_{U}|\Psi\rangle_{HP}\right)_{{\rm conn}}+\mathcal{O}(U^{3})\]
 where $(...)_{{\rm conn}}$ indicates a sum over connected diagrams
and $H_{0}$ denotes the unperturbed Hamiltonian. The first order
contribution is obtained rewriting the number operators in momentum
space, thus obtaining \begin{equation}
_{HP}\langle\Psi|H_{U}|\Psi\rangle_{HP}=LUn_{\uparrow}n_{\downarrow}\label{eq:EE1}\end{equation}
 while the second order contribution can be computed evaluating Goldstone
diagrams:\begin{equation}
\mathcal{E}_{HP}^{(2)}=-\frac{U^{2}}{t(2\pi)^{3}}\frac{\vartheta(k_{F\uparrow},k_{\downarrow},a)}{(1+z)}\label{eq:EE2}\end{equation}
 \[
\vartheta(k_{F\uparrow},k_{\downarrow},a)\equiv\int_{0}^{\frac{\pi}{2}}\frac{dq}{\sin q}\int_{-q}^{q}\; dk\times\]
 \[
\times\int_{-q}^{q}\frac{dk'}{\sin(k+k_{F\uparrow})+a\sin(k'+k_{F\downarrow})}.\]
 Integrating the previous expression numerically, we can give a quantitative
estimate of the phase boundary near the FK limit. Furthermore, by
comparing ${\cal E}_{HP}$ up to second order (plotted in Fig. \ref{fig:edensity})
with previous numerical results \cite{F2007}, we have a good check
that at half filling the variational ground state (\ref{eq:PsiE})
is a correct description of the system even at finite $U/t_{\uparrow}\leq3$.

\subsection{Phase boundaries}

The weak-coupling phase diagram is generally characterized by two
types of phase transition: one between the HP state and the PS region,
and the other one between different types of spatially separated states.

\begin{figure}
\includegraphics[height=4cm]{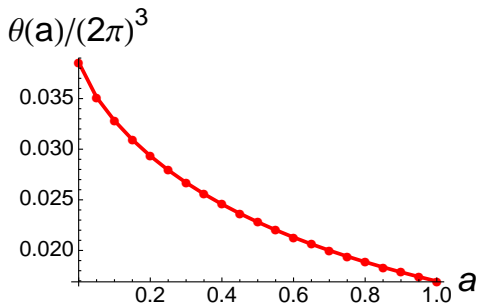}

\caption{Second-order perturbative contribution at half filling $\vartheta(\pi/2,\pi/2,a)/(2\pi)^{3}$
for different values of $a$; solid line is a guide for the eye.\label{fig:pdweak2}}

\end{figure}

Combining Eqs.(\ref{eq:eps}), (\ref{eq:ET}), (\ref{eq:EE1}) and
(\ref{eq:EE2}), the first mentioned phase transition line is determined
by the equation: \[
2\sin\left(\frac{\pi n}{2}\right)-(1+z)\frac{2-n}{2}\sin\left(\frac{\pi n}{2-n}\right)=\]
 \begin{equation}
\frac{\pi}{2}u\left[\frac{n^{2}}{4}-u\frac{\vartheta(k_{F\uparrow},k_{F\uparrow},a)}{(2\pi)^{3}(1+z)}\right]\label{eq:PT}\end{equation}
 Table \ref{theta} shows how the correlated energy factor depends
on $k_{F\uparrow}$ and $z=(1-a)/(1+a)$. In general, the larger the
asymmetry $z$ is (the smaller is $a)$ the larger is $\vartheta$.
An illustrative plot at half-filling is presented in Fig. \ref{fig:pdweak2}.
Furthermore $\vartheta$ approaches 0 in the low-density limit and
grows with the filling.

\begin{table}[htbp]
 \begin{tabular}{|c|c|c|c|}
\hline 
$a\backslash k_{F\downarrow}=\pi n/2$  & $\pi/3$  & $\pi/4$  & $\pi/6$\tabularnewline
\hline 
0.01  & 0.028  & 0.022  & 0.013\tabularnewline
\hline 
0.05  & 0.027  & 0.021  & 0.012\tabularnewline
\hline 
0.1  & 0.026  & 0.02  & 0.012\tabularnewline
\hline 
0.15  & 0.025  & 0.019  & 0.012\tabularnewline
\hline 
0.2  & 0.024  & 0.018  & 0.011\tabularnewline
\hline
\end{tabular}

\caption{Numerical values for $\vartheta(k_{F\uparrow},k_{F\uparrow},a)/(2\pi)^{3}$
below half filling. Notice that the term $(1+z)$ in the denominator
of Eq. (\ref{eq:EE2}) has not been included in the definition of
$\vartheta$ so to have a more direct comparison with ref. \cite{MV1989}.\label{theta}}

\end{table}

We will consider first the case when the density is medium-high: $n_{\uparrow}=n_{\downarrow}=n/2>0.25$,
when a transition from HP to TSS state should always take place, being
$\nu^{*}>n/2$ unfavored. By inspecting Eq. (\ref{eq:PT}) it turns
out that the ground state in the weak coupling limit is always homogeneous,
even for large $z$ ($a<0.2$, i.e. $z>2/3$). This fact agrees with
previous numerical results \cite{F2007} showing that the TSS phase
is present only for $U/t_{\uparrow}\gtrsim2.5$. Even close to quarter
filling, the phase transition in the FK limit is predicted at $U/t_{\uparrow}=1.15$,
which is beyond the weak coupling regime we are considering in this
section.

\begin{figure}
\includegraphics[height=7cm]{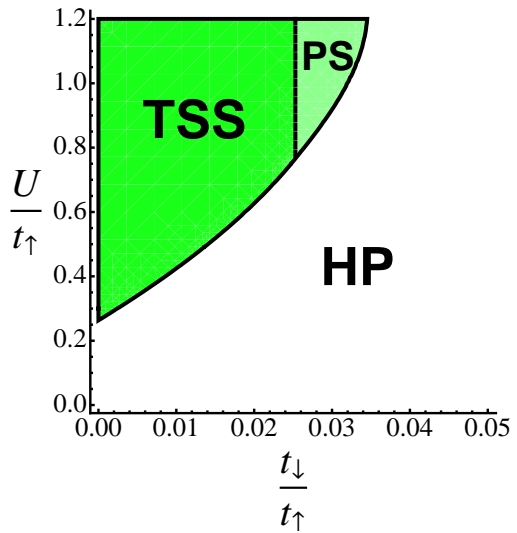}

\caption{{\small Weak-coupling phase diagram with $n=1/3$: white region represents
the homogeneous state (HP), green regions are different phase separated
states. The phase transition lines are computed using Eq. (\ref{eq:znuunb})
(dashed) and Eq. (\ref{eq:PT}) (solid).} In this section we have
adopted the parameters $U/t_{\uparrow}=u/(1+z)$ and $a=t_{\downarrow}/t_{\uparrow}=(1-z)/(1+z)$
instead of $u$ and $z$ to facilitate the comparison with previous
results in the weak-coupling region.$ $\label{fig:pdweak}}

\end{figure}

Let us examine now the low density regime: $n_{\uparrow}=n_{\downarrow}=n/2<0.25$,
where the TSS might be the ground state only at very small $a$. Now
we have to determine the transition point from HP to TSS state in
the FK limit ($a$ very small) as well as to explore the possibility
of a transition to a PS state with $\nu>n_{\downarrow}$ for larger
values of $a$. We will examine first the $n=1/3$ case as an example.
In the highly asymmetric regime ($a<0.01$, i.e. $z>0.98$), we again
find a transition from a HP to a TSS state which happens for values
of $U(n=1/3)/t_{\uparrow}\sim0.3$. As explained in the previous section,
using Eqs. (\ref{eq:eps}) and (\ref{eq:znuunb}) we can determine
the phase transition line between TSS and a PS state with $\nu\neq n_{\downarrow}$.
This transition occurs for $t_{\downarrow}(n=1/3)/t_{\uparrow}\sim0.025$,
a value which turns out to be independent of $U$ since no contributions
from the correlation energy are present. These transition points are
in good accordance with the numerical results reported in \cite{F2007},
and allow us to complete a general weak coupling phase diagram in
the highly asymmetric regime for $n=1/3$, which is shown in Fig.
\ref{fig:pdweak}. At smaller densities, the critical value $U(n)/t_{\uparrow}$
at which one finds the transition from the HP to the TSS state becomes
smaller. For example $U(n=1/6)/t_{\uparrow}\sim0.06$, while $U(n=1/12)/t_{\uparrow}\lesssim10^{-3}$.
We notice that it is impossible, within our perturbative approach
(see Eq. (\ref{eq:PT})), to find for this coefficient a value equal
to zero: its value decreases as $n$ becomes smaller but stays always
finite, going to zero smoothly as $n$ tends to zero. In addition,
the TSS is always unstable with respect to a PS one, the transition
now appearing at lower asymmetries ($t_{\downarrow}(n=1/6)/t_{\uparrow}\sim2.5\times10^{-3}$,
$t_{\downarrow}(n=1/12)/t_{\uparrow}\sim2.7\times10^{-4}$), whose
values are still essentially insensitive to $U$.

\section{Strong-coupling limit\label{sec:s-c}}

In this section we analyze the case of strong repulsive onsite interaction
between fermions, corresponding to $t_{\sigma}\ll U$, a regime which
is of particular interest for the experimental realization of the
symmetric ($z=0$) model in a cold atom system \cite{Jetal2008},
in which a Mott-insulator phase at half filling was found. One of
the questions left open by bosonization is what happens to the HP-PS
transition curve close to $z=0$. From the phase diagrams in refs.
\cite{CHG2005} and \cite{M2007} it is not clear whether it approaches
a finite value when $z\to0$ or, conversely, if there is a characteristic
value of $z\neq0$ at which it diverges, as indicated also by some
data on short sizes in ref. \cite{GFL2007} (Sec. IV therein). To
study this regime, we will construct an effective Hamiltonian that
is able to describe the AHM when the parameter $t$ can be considered
as a small perturbation with respect to $U$, for a generic filling,
by using the method of the flow equations, developed by Wegner \cite{W1994}
and applied to the HM in \cite{S1997}. The advantages of using such
a technique are extensively described in \cite{S1997}. We only remark
here that it yields a very general procedure which, in a recursive
way, allows to find an effective Hamiltonian at any order of perturbation,
for arbitrary fillings and geometries.

We start by decomposing the fermionic Hilbert space of the model into
the subspaces $\mathcal{H}_{k}$ with exactly $k$ fermionic pairs
(double occupancies): $\mathcal{H}=\oplus_{k=0}^{N/2}\mathcal{H}_{k}$.
The projectors $P_{k}$ on these subspaces are defined via the generating
function \[
\sum_{k=0}^{N}P_{k}x^{k}=\prod_{i\in L}\left[1-(1-x)n_{i\uparrow}n_{i\downarrow}\right].\]
 The kinetic energy term for the spin $\sigma$ fermions, $T_{\sigma}=\sum_{<ij>}c_{i\sigma}^{\dagger}c_{j\sigma}$,
can be decomposed into three parts $T_{\sigma}=T_{0\sigma}+T_{+1\sigma}+T_{-1\sigma}$,
which change the number of pairs by $\delta k=m=0,+1,-1$. In other
words: $T_{m\sigma}=\sum_{k=0}^{N}P_{k+m}T_{\sigma}P_{k}$. In these
expressions, we have introduced the sum over $<ij>$ which denotes
nearest-neighbors sites $i$ and $j$ (with the couples $ij$ and
$ji$ counted once each) since the procedure we are going to discuss
is generalizable to any dimension. More explicitly: \begin{eqnarray*}
 &  & T_{0\sigma}=\sum_{<ij>}\left[n_{i\bar{\sigma}}c_{i\sigma}^{\dagger}c_{j\sigma}n_{j\bar{\sigma}}+(1-n_{i\bar{\sigma}})c_{i\sigma}^{\dagger}c_{j\sigma}(1-n_{j\bar{\sigma}})\right]\\
 &  & T_{+1\sigma}=\sum_{<ij>}n_{i\bar{\sigma}}c_{i\sigma}^{\dagger}c_{j\sigma}(1-n_{j\bar{\sigma}})\\
 &  & T_{-1\sigma}=\sum_{<ij>}(1-n_{i\bar{\sigma}})c_{i\sigma}^{\dagger}c_{j\sigma}n_{j\bar{\sigma}}\end{eqnarray*}
 It is not difficult to verify that $\left[N_{d},T_{m\sigma}\right]=mT_{m\sigma}$,
where $N_{d}=\sum_{i}n_{i\uparrow}n_{i\downarrow}$, reflecting the
transition from the Hilbert space $\mathcal{H}_{k}$ to $\mathcal{H}_{k+m}$.
To discuss higher-order interactions terms, it is useful to introduce
products of hopping operators, $T_{\sigma_{1}\dots\sigma_{k}}^{(k)}(\mathbf{m})=T_{m_{1}\sigma_{1}}...T_{m_{k}\sigma_{k}}$
with the index vector $\mathbf{m}=(m_{1},m_{2},...,m_{k})$. It is
found that the commutator of such an operator product with $N_{d}$
involves the total weight of the product, $M(\textbf{m})=\sum_{i=1}^{k}m_{i}$,
and generally reads $\left[N_{d},T_{\sigma_{1}\dots\sigma_{k}}^{(k)}(\mathbf{m})\right]=M(\mathbf{m})T_{\sigma_{1}\dots\sigma_{k}}^{(k)}(\textbf{m})$.
We want now to find an effective Hamiltonian which does not mix the
different Hilbert space sectors $\mathcal{H}_{k}$, i.e. which conserves
the total number of local pairs, thus suited to study physical properties
at energy and temperature scales which are well below the Hubbard
energy $|U|$. To do so, we consider a continuous unitary transformation
which allows to remove interactions with non-vanishing overlap between
different Hilbert space sectors. Thus, the transformed Hamiltonian
depends on a continuous flow parameter $l$:\[
H(l)=-\sum_{\sigma}t_{\sigma}\Theta_{\sigma}(l)+UN_{d}\]
 where the generalized kinetic energy term $\Theta_{\sigma}(l)$ contains
all order interactions which are generated by the transformation:
\begin{equation}
\Theta_{\sigma}(l)=\sum_{k=1}^{\infty}\frac{t_{\sigma}^{k-1}}{U^{k-1}}\sum_{\{\mathbf{m}\}}F_{\sigma_{1}...\sigma_{k}}^{(k)}(l;\mathbf{m})T_{\sigma_{1}...\sigma_{k}}^{(k)}(\mathbf{m})\label{c:1}\end{equation}
 Here $F_{\sigma_{1}...\sigma_{k}}^{(k)}$ denote suitable coupling
functions that have to be determined by asking that the unitary transformation
cancels all terms that do not conserve the number of local pairs.
The flow equations for these coupling functions follow from the equation
of the flow for the Hamiltonian \cite{W1994}: \begin{equation}
\frac{dH(l)}{dl}=\left[\eta_{\sigma}(l),H(l)\right]\label{fe}\end{equation}
 which has been written here by using the (antihermitean) generator
of the transformation\[
\eta_{\sigma}(l)=\frac{t_{\sigma}}{U}\left[V,\Theta_{\sigma}(l)\right]=\]
 \[
=\sum_{k=1}^{\infty}\frac{t_{\sigma}^{k}}{U^{k}}\sum_{\{\mathbf{m}\}}M(\mathbf{m})F_{\sigma_{1}...\sigma_{k}}^{(k)}(l;\mathbf{m})T_{\sigma_{1}...\sigma_{k}}^{(k)}(\mathbf{m})\]

Now, after imposing both the initial conditions ($F_{\sigma_{1}}^{(1)}(0;m)=1$
and $F_{\sigma_{1}\dots\sigma_{k}}^{(k)}(0;\textbf{m})=0$ for $k>1$)
and the symmetries related to hermiticity and particle-hole transformation
$c_{i\sigma}^{\dagger}\rightarrow c_{i\sigma}$, which reverses the
sign of the hopping term, one can recast the original flow equation
(\ref{fe}) in a recursive set of coupled nonlinear differential equations.
From these equations it is easy to see that all the terms which connect
different sectors of the Hilbert space vanish in the limit $l\rightarrow\infty$.
At the second order we find that the effective Hamiltonian reads \[
H_{s-c}=-t_{\uparrow}\sum_{<ij>}\Big[n_{i\downarrow}c_{i\uparrow}^{\dagger}c_{j\uparrow}n_{j\downarrow}+(1-n_{i\downarrow})c_{i\uparrow}^{\dagger}c_{j\uparrow}(1-n_{j\downarrow})\Big]\]
 \[
-t_{\downarrow}\sum_{<ij>}\Big[n_{i\uparrow}c_{i\downarrow}^{\dagger}c_{j\downarrow}n_{j\uparrow}+(1-n_{i\uparrow})c_{i\downarrow}^{\dagger}c_{j\downarrow}(1-n_{j\uparrow})\Big]\]
 \begin{equation}
+J\sum_{<ij>}\Big[S_{i}^{x}S_{j}^{x}+S_{i}^{y}S_{j}^{y}+\Delta(S_{i}^{z}S_{j}^{z}-n_{i}n_{j}/4)\Big]+UN_{d}\label{eq:Hs-c}\end{equation}
 where $S_{i}^{x,y,z}$ are the spin operators at site $i$, $J=2t_{\uparrow}t_{\downarrow}/U$
and $\Delta=(t_{\uparrow}^{2}+t_{\downarrow}^{2})/2t_{\uparrow}t_{\downarrow}\ge1$.
At half filling, $n=1$, we get, according to the Takahasi's theorem
\cite{T1977}, that the terms corresponding to the odd orders of
the expansion (in the our case the first two lines) vanish and we
find the same Hamiltonian obtained in \cite{FDL1995} representing
a spin chain with an anisotropy term $\Delta\ge1$ along the $z-$axis
(XXZ chain): the spin excitations are gapped and the spin-spin correlators
decay exponentially with the distance. Also, in the limit $t_{\uparrow}=t_{\downarrow}$
(symmetric HM) the anisotropy term $\Delta$ becomes 1 and we find
the well-known Heisenberg Hamiltonian (XXX chain), as it should be.

It is well known \cite{FDL1995} that for $z>0$ the system is in
the Néel-like phase, with non-vanishing charge and spin gap and true
long-range order. Here we are interested in examining the two limiting
cases $z=1$ (FK model) and $z=0$ (Hubbard model) to study, more
precisely, $(i)$ the phase appearing in the Hubbard model when $U\rightarrow\infty$,
i.e. the so called spinless fermions phase (SF), where the orientation
of the spins loses its relevance since the doubly occupied sites are
strictly forbidden, and $(ii)$ the state predicted in the FK model
where the two fermionic species are demixed.

\subsection{Spinless fermions }

\begin{figure}
\includegraphics[height=7cm]{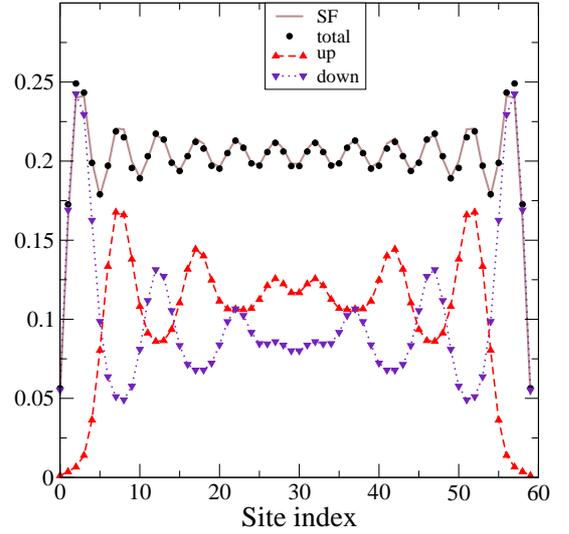}

\caption{Spatial density profiles for an open chain with $z=0.1$, filling
$n=0.2$ and $u=100$. The continuous curve labelled SF is the local
density for an equivalent chain of spinless fermions (i.e. Eq. (\ref{eq:dSF})
with $\ell=L$ and $p=nL)$, showing a very good agreement with the
spinful numerical data.\label{fig:z01_vs_SF}}

\end{figure}

The SF state $\vert\Psi\rangle_{SF}$ of the Hubbard model in the
limit $U\to\infty$ at filling $n=N/L\le1$ and equally populated
species $n_{\uparrow}=n_{\downarrow}=n/2$ is rotationally invariant
and invariant under the up-down exchange. In this case, the expectation
value of the hopping terms of the Hamiltonian (\ref{eq:Hs-c}) reads
as follows\[
_{SF}\langle\Psi\vert[-t\sum_{j,\sigma}(c_{j\sigma}^{\dagger}c_{j+1\sigma}+{\rm h.c.})-tz\sum_{j}(c_{j\uparrow}^{\dagger}c_{j+1\uparrow}+{\rm h.c.})\]
 \[
+tz\sum_{j}(c_{j\downarrow}^{\dagger}c_{j+1\downarrow}+{\rm h.c.})]\vert\Psi\rangle_{SF}=-\frac{2t}{\pi}\sin(\pi n)\]
 As for the $J$-terms, we can borrow directly its expression from
(A3) of \cite{OS1990}: \begin{equation}
\frac{_{SF}\langle\Psi\vert\sum_{j}n_{j}n_{j+1}\vert\Psi\rangle_{SF}}{L}=n^{2}-\frac{\sin^{2}(\pi n)}{\pi^{2}}\label{eq:nnSF}\end{equation}
 Moreover, since $\vert\Psi\rangle_{SF}$ is SU(2)-invariant, we can
write \begin{equation}
\frac{_{SF}\langle\Psi\vert\sum_{j}S_{j}^{\alpha}S_{j+1}^{\alpha}\vert\Psi\rangle_{SF}}{L}=\frac{1}{3}\frac{_{SF}\langle\Psi\vert\sum_{j}\vec{S}_{j}\cdot\vec{S}_{j+1}\vert\Psi\rangle_{SF}}{L}\label{eq:SU2SF}\end{equation}
 for $\alpha=x,y,z$ even if the $J$-part in the strong-coupling
Hamiltonian is anisotropic, where now, from (A3) and (A4) of of \cite{OS1990},
we find\[
\frac{_{SF}\langle\Psi\vert\sum_{j}\vec{S}_{j}\cdot\vec{S}_{j+1}\vert\Psi\rangle_{SF}}{L}=\left(\frac{1}{4}-\ln2\right)\left[n^{2}-\frac{\sin^{2}(\pi n)}{\pi^{2}}\right].\]
 Collecting all the pieces it is easy to see that\[
{\cal {\cal E}}_{SF}=\frac{_{SF}\langle\Psi\vert H_{s-c}\vert\Psi\rangle_{SF}}{L}=-\frac{2t}{\pi}\sin(\pi n)+\]
 \[
+\frac{4t^{2}(1-z^{2})}{U}\left[n^{2}-\frac{\sin^{2}(\pi n)}{\pi^{2}}\right]\left[\frac{2+\Delta}{3}\left(-\ln2+\frac{1}{4}\right)-\frac{1}{4}\right]\]
 As an example, in Fig. \ref{fig:z01_vs_SF} we plot the local densities
of fermions obtained numerically on a chain with $L=60$ and open
boundary conditions (OBC), $z=0.1$, $n=0.2$ and $u=100$. The two
species tend to occupy alternate regions but the fraction of double
occupation is still significant. The comparison with the total density
profile for spinless fermions at the same equivalent filling shows
that the SF state is a good description of the ground state in this
case. We have verified that the same happens if the filling is increased
up to $n=0.9$, the other parameters being unchanged. On the contrary,
if we still fix $u=100$, $n=0.2$ but increase the anisotropy to
$z=0.3$, appreciable differences in the density profiles start to
appear.

\subsection{Spatially separated states}

In the strong-coupling approach we can actually formulate a slightly
more general form of the PS state with respect to that of Sec. \ref{sub:twfw-c}
which, however, leads formally to the same analytical expressions.
Let us consider a sequence of $M$ contiguous intervals $I_{j\uparrow}$,
$I_{j\downarrow}$ ($j=1,\dots,M$) and a state $\vert\Psi\rangle_{PS}$
in which the up and down spins are separated in the sense that there
are no doubly occupied sites, $I_{1\uparrow}$ contains only up spins,
$I_{1\downarrow}$ contains only down spins, then again $I_{2\uparrow}$
with up spins and so on. Let $L_{j\sigma}$ and $N_{j\sigma}$, respectively,
the number of sites and the number of electrons in the interval $I_{j\sigma}$.
We then make the further strong assumption that each up or down interval,
irrespective of its length, is equally filled meaning that $\bar{n}_{\uparrow}=N_{j\uparrow}/L_{j\uparrow}$
and $\bar{n}_{\downarrow}=N_{j\downarrow}/L_{j\downarrow}$ do not
depend on $j$. Now, if\[
\nu L=\sum_{j=1}^{M}L_{j\downarrow}\;,\;\;(1-\nu)L=\sum_{j=1}^{M}L_{j\uparrow}\]
 are the total lengths associated with the motion of down and up spins
we have $N_{\uparrow}=\sum_{j=}^{M}N_{j\uparrow}=\sum_{j=1}^{M}\bar{n}_{\uparrow}L_{j\uparrow}=\bar{n}_{\uparrow}L(1-\nu)$
and similarly $N_{\downarrow}=\bar{n}_{\downarrow}L\nu$ and so, for
equally populated species $\bar{n}_{\uparrow}=n/[2(1-\nu)]$ and $\bar{n}_{\downarrow}=n/(2\nu)$.

We first consider the thermodynamic limit in the case in which the
interface points (which are $2M$ in number) do not contribute to
the bulk energy density ($\lim_{L\to\infty}2M/L=0$) and, at the same
time, each interval is extensively large ( $\lim_{L\to\infty}L_{j\sigma}/L>0$),
so that for every interval we can use the formula for the kinetic
energy density of free fermions without worrying about finite-size
and/or boundary effects:\[
\frac{_{PS}\langle\Psi\vert T_{s-c}\vert\Psi\rangle_{PS}}{L}=\frac{\sum_{j=1}^{M}L_{j\uparrow}}{L}\left[-\frac{2t_{\uparrow}}{\pi}\sin\left(\pi\frac{n}{2(1-\nu)}\right)\right]\]
 \[
+\frac{\sum_{j=1}^{M}L_{j\downarrow}}{L}\left[-\frac{2t_{\downarrow}}{\pi}\sin\left(\pi\frac{n}{2\nu}\right)\right]\]
 \[
=-\frac{2t_{\uparrow}}{\pi}(1-\nu)\sin\left(\pi\frac{n}{2(1-\nu)}\right)-\frac{2t_{\downarrow}}{\pi}\nu\sin\left(\pi\frac{n}{2\nu}\right).\]
 Note that we do not necessarily require that the heavy (down) species
is fully compressed, meaning $\nu=n/2$. The value for $\nu$ will
be determined variationally in order to give the smallest possible
energy at a fixed $z$, exactly as done in Sec. \ref{sub:ground stateew-c}.
Let us now calculate the energy of such a state.

\begin{figure}
\includegraphics[height=7cm]{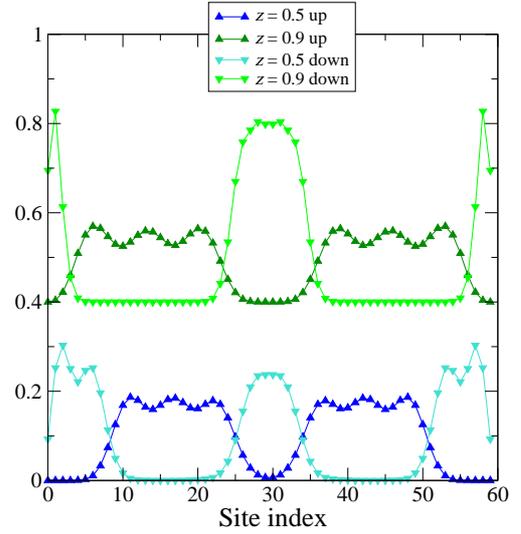}

\caption{Spatial density profiles for open chains with $z=0.5$ and $z=0.9$
(offset by $+0.4$ for clarity), filling $n=0.2$ and $u=100$.\label{fig:z05_09}}

\end{figure}

As far as the $J$-term is concerned we first note that the transverse
part is vanishing. In fact both $S_{j}^{x}$ and $S_{j}^{y}$ are
composed by spin-flip terms like $c_{j\sigma}^{\dagger}c_{j\bar{\sigma}}$
but each interval contains spins of only one specie. Next, all the
$\Delta$-term can be rewritten as\[
\Delta\left(S_{j}^{z}S_{j+1}^{z}-\frac{n_{j}n_{j+1}}{4}\right)=\frac{\Delta}{4}[(n_{j\uparrow}-n_{j\downarrow})(n_{j+1\uparrow}-n_{j+1\downarrow})\]
 \begin{equation}
-(n_{j\uparrow}+n_{j\downarrow})((n_{j+1\uparrow}+n_{j+1\downarrow})]=-\frac{\Delta}{2}(n_{j\uparrow}n_{j+1\downarrow}+n_{j\downarrow}n_{j+1\uparrow}).\label{eq:surfterm}\end{equation}
When the expectation value on $\vert\Psi\rangle_{PS}$ is taken, the
up and down parts factorize and there can be non-vanishing contributions
only when $j$ and $j+1$ are at an interface between two intervals
carrying opposite spins. If, as assumed above, the number of interface
points does not grow as $L$ we can neglect these contributions in
the limit $L\to\infty$. Therefore$ $ the energy density of the PS
state reads\[
{\cal E}_{PS}(\nu)=\frac{_{PS}\langle\Psi\vert H_{s-c}\vert\Psi\rangle_{PS}}{L}\]
 \[
=-\frac{2t_{\uparrow}}{\pi}(1-\nu)\sin\left(\pi\frac{n}{2(1-\nu)}\right)-\frac{2t_{\downarrow}}{\pi}\nu\sin\left(\pi\frac{n}{2\nu}\right).\]
As anticipated, this expression coincides with Eq. (\ref{eq:eps})
in the balanced case $n_{\uparrow}=n_{\downarrow}=n/2$. In Fig. \ref{fig:z05_09}
we present two examples of the spatial density profile for large $u$
and intermediate/large $z$, from which the spatial separation of
the two species can be clearly inferred. Note also that the light
specie occupies regions with a non-vanishing fraction of empty sites
and an oscillating density profile $\langle n_{j\uparrow}\rangle$
is seen. Nonetheless the local density in the intervals occupied by
the heavy fermions does not reach 1, so in these case we do not have
a TSS as instead, for example, in Fig. 3 of ref. \cite{S-VFF2007}
valid for $u=20$, $z=2/3$, filling $n=0.8$ on 40 sites (reproduced
in our calculations but not shown here).

\subsection{PS': Extensive number of intervals $M=\alpha L$\label{sub:MproptoL}}

In order to treat also the case in which the number of interfaces
scales as a finite fraction $\alpha$ of the total number of sites,
we will assume that all the intervals hosting up spins are equally
long and equally filled: $L_{j\uparrow}=\ell_{\uparrow}$, $N_{j\uparrow}=p_{\uparrow}$
$\forall j=1,\dots,M$ so that $\ell_{\uparrow}=(1-\nu)/\alpha$ and
$p_{\uparrow}=n/(2\alpha)$; similarly for the intervals with down
spins $\ell_{\downarrow}=\nu/\alpha$ and $p_{\downarrow}=n/(2\alpha)=p_{\uparrow}$.
Note that, for equally populated species, necessarily the finite number
of electrons in each interval is the same for up or down spins, while
the finite lengths are in general different. The energy density of
this type of phase separated state will therefore have the form\[
{\cal E}_{PS'}=\frac{M}{L}(\kappa_{\uparrow}+\kappa_{\downarrow})-\frac{J\Delta}{L}\sum_{j=1}^{L}(\langle n_{j\uparrow}\rangle\langle n_{j+1\downarrow}\rangle+\langle n_{j\downarrow}\rangle\langle n_{j+1\uparrow}\rangle)\]
 where $\kappa_{\uparrow,\downarrow}$ are the kinetic energies of
$p=p_{\uparrow}=p_{\downarrow}$ up or down fermions on intervals
of length $\ell_{\uparrow,\downarrow}$ with OBC, while the $\Delta$-term
comes from Eq. (\ref{eq:surfterm}) and now cannot be neglected. The
on-site terms $\langle n_{j\sigma}\rangle$ also have to be evaluated
in the same fashion and will be localized at the left or right end
of the intervals (with equal values). Let us denote them by $\eta_{\sigma}$;
we have a contribution $\langle n_{{\rm right}\uparrow}\rangle\langle n_{{\rm left}\downarrow}\rangle+\langle n_{{\rm right}\downarrow}\rangle\langle n_{{\rm left}\uparrow}\rangle=2\eta_{\uparrow}\eta_{\downarrow}$
for each of the $M$ pairs of up+down intervals so that\[
{\cal E}_{PS'}=\alpha(\kappa_{\uparrow}+\kappa_{\downarrow}-2J\Delta\eta_{\uparrow}\eta_{\downarrow}).\]

The calculation of the kinetic energy and of the surface density for
an effective open chain of $\ell$ sites with $p$ free fermions is
given in the appendix, leading to:\[
{\cal E}_{PS'}=\alpha t\{(1+z)\left\{ 1-\frac{\sin\left[\frac{\pi(2p+1)}{2(\ell_{\uparrow}+1)}\right]}{\sin\left[\frac{\pi}{2(\ell_{\uparrow}+1)}\right]}\right\} \]
 \[
+(1-z)\left\{ 1-\frac{\sin\left[\frac{\pi(2p+1)}{2(\ell_{\downarrow}+1)}\right]}{\sin\left[\frac{\pi}{2(\ell_{\downarrow}+1)}\right]}\right\} \]
 \[
-\frac{1+z^{2}}{u}\frac{\left\{ 2p+1-\frac{\sin\left[\frac{\pi(2p+1)}{\ell_{\uparrow}+1}\right]}{\sin\left(\frac{\pi}{\ell_{\uparrow}+1}\right)}\right\} \left\{ 2p+1-\frac{\sin\left[\frac{\pi(2p+1)}{\ell_{\downarrow}+1}\right]}{\sin\left(\frac{\pi}{\ell_{\downarrow}+1}\right)}\right\} }{(\ell_{\uparrow}+1)(\ell_{\downarrow}+1)}\}.\]
 The conditions $p\le\ell_{\uparrow}$ and $p\le\ell_{\downarrow}$
define the range of $\nu\in[n/2,1-n/2]$, while the conditions $p\ge1$,
$\ell_{\uparrow}\ge1$ and $\ell_{\downarrow}\ge1$ imply $\alpha\le\min(n/2,\nu,1-\nu)=n/2$.
Once this expression is minimized by suitable values of $\alpha$
and $\nu$ in this range we should, at least, check if the resulting
energy density is smaller than the optimal energy density ${\cal E}_{PS}$
determined above for the same values of $z$, $n$ and, now, also
$u$.

Finally, we mention that we have also tried to enlarge the set of
trial/variational states by including the homogeneous one (defined
in Sec. \ref{sec:w-c}), which is the correct ground state in the
limit $U=0$ for all $z$. However we have verified that this additional
state, for the fillings we have considered, could become relevant
only when $u\lesssim1$, outside the domain of validity of the strong-coupling
approach. Therefore, for the sake of compactness, we do not report
these results here.

\subsection{Phase boundaries}

From the condition ${\cal E}_{PS}<{\cal E}_{SF}$ we get

\[
-\frac{2(1+z)}{\pi}(1-\nu^{*})\sin\left(\pi\frac{n}{2(1-\nu^{*})}\right)\]
 \[
-\frac{2(1-z)}{\pi}\nu^{*}\sin\left(\pi\frac{n}{2\nu^{*}}\right)<-\frac{2}{\pi}\sin(\pi n)\]
 \[
+\frac{4(1-z^{2})}{u}\left[n^{2}-\frac{\sin^{2}(\pi n)}{\pi^{2}}\right]\left[\frac{2+\Delta}{3}\left(-\ln2+\frac{1}{4}\right)-\frac{1}{4}\right]\]
 that is $u<u_{PS}(z)$ for $d(n,z)>0$ or $u>u_{PS}(z)$ otherwise,
having defined \[
d(n,z)=\sin(\pi n)-(1+z)(1-\nu^{*})\sin\left[\pi\frac{n}{2(1-\nu^{*})}\right]\]
 \[
-(1-z)\nu^{*}\sin\left(\pi\frac{n}{2\nu^{*}}\right)\]
 and\[
u_{PS}(n,z)=2\pi(1-z^{2})\frac{\left[n^{2}-\frac{\sin^{2}(\pi n)}{\pi^{2}}\right]\left[\frac{2+\Delta}{3}\left(-\ln2+\frac{1}{4}\right)-\frac{1}{4}\right]}{d(n,z)}\]

\begin{figure}
\includegraphics[height=7cm]{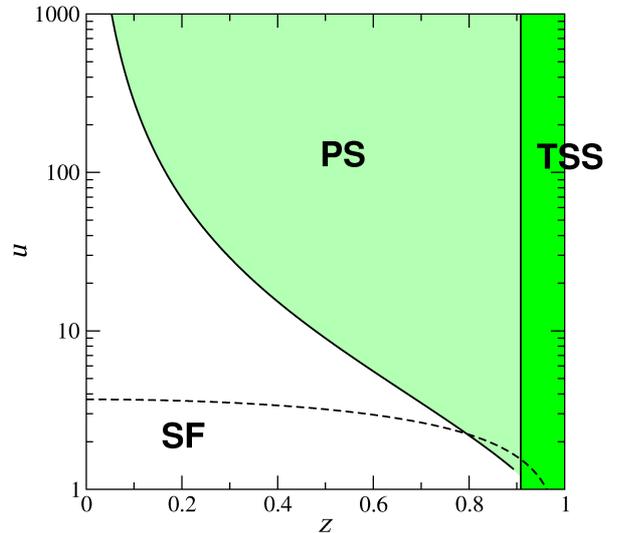}

\caption{Transition lines in the $(z,u)$-plane at filling $n=0.4$ indicating
how the ground state changes from SF to PS. In the separated regime
above the transition lines, the state can be either PS (left) or TSS
(right) (see Sec. \ref{sub:ground stateew-c}) and the edge between
the two is marked by a vertical line (analogously to Fig. 3 of ref.
\cite{F2007}).$ $ The dashed line corresponds to the bosonization
prediction.\label{fig:pd04}}

\end{figure}

\begin{figure}
\includegraphics[height=7cm]{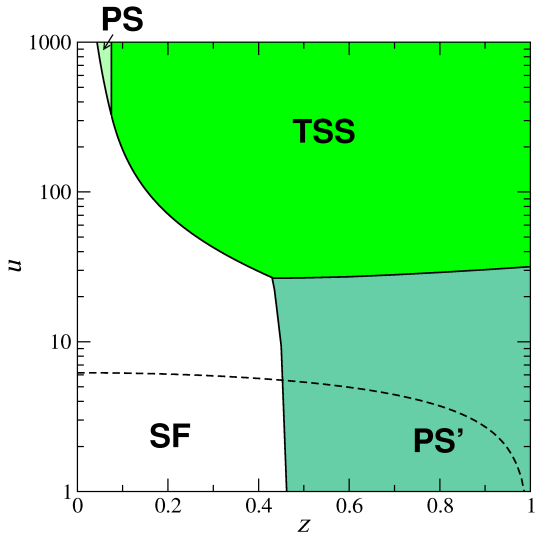}

\caption{Same as Fig. \ref{fig:pd04} but with filling $n=0.9$. Now also the
PS' (see text) state is relevant for moderate values of $u$ and the
corresponding transition lines with SF and PS are marked.$ $\label{fig:pd09}}

\end{figure}

Now we can draw a phase diagram in the $(z,u)$-plane for a fixed
value of the total filling $n$ by indicating the regions where the
PS or the SF state has the lower energy. We show two examples (for
$n=0.4$ and $0.9$) in Figs. \ref{fig:pd04} and \ref{fig:pd09},
respectively. We have analyzed in detail also the phase diagram for
$n=0.6$ (not shown), that turns out be qualitatively similar to the
one for $n=0.4$. Having in mind that bosonization could be considered
quantitatively reliable only for small values of the interaction,
we have reported in the figures (dashed lines) the curves of Wentzel-Bardeen
instability \cite{M2007} where the velocity of one of the bosonization
modes vanishes thereby indicating phase separation. In addition, we
have considered two typical cases of PS at $n=0.9$ $(L=20,40,60)$,
namely those at $u=100$ for $z=0.1$ and at $u=5$ for $z=0.9$.
In both cases, the charge structure factor $S(q)$ (as defined in
Sec. \ref{sec:uneg}) displays a divergence for $q\to0$ typical of
PS states \cite{SBC2004} and a peak at $q=4k_{F}({\rm mod}2\pi)=2\pi n({\rm mod}2\pi)=2\pi(1-n)$.

In the strong-coupling approach the most interesting thing to understand
seems to be the divergence of the transition line separating PS from
SF behaviour at small $z$ and large $u$. We have verified that when
$\nu=\nu^{*}(z)$ is inserted into $d(n,z)$ the denominator appearing
in $u_{PS}(n,z)$ is always negative. In order to estimate $u_{PS}(n,z)$
analytically we set $\nu^{*}(z)=1/2-z\nu_{1}-z^{2}\nu_{2}+O(z^{3})$
with $\nu_{1,2}>0$ (see Fig. \ref{fig:zbnus}) and expand $d(n,z)$
for $z\to0$\[
d(n,z)=-2\nu_{1}[f(\pi n)-\pi^{2}n^{2}\nu_{1}\sin(\pi n)]z^{2}+O(z^{3})\]
 so that at leading order\begin{equation}
u_{PS}(n,z)=\frac{1}{z^{2}}\frac{\pi\ln2\left[n^{2}-\frac{\sin^{2}(\pi n)}{\pi^{2}}\right][1+O(z)]}{\nu_{1}(n)[f(\pi n)-\pi^{2}n^{2}\nu_{1}(n)\sin(\pi n)]}\label{eq:PSlargeu}\end{equation}
 with $f(x)=\sin x-x\cos x$ as before. As far $\nu_{1}(n)$ as is
concerned, by inserting $\nu=1/2-z\nu_{1}+O(z^{2})$ in Eq. (\ref{eq:znu})
and solving at first order in $z$ we get $\nu_{1}(n)=f(\pi n)/[f'(\pi n)2\pi n]$.
In summary, our analytical approach predicts that there is no finite
value of $z$ below which PS disappears; by moving to a sufficiently
large repulsive coupling it is always possible to induce a PS at arbitrarily
small anisotropy.

\subsection{Inclusion of phase separated states with an infinite number of interfaces}

We have compared the energy density of the PS state at given values
of $n$ and $z$ with the corresponding value for the PS' state discussed
in Sec. \ref{sub:MproptoL}. At small filling, say $n=0.1$, the two
variational solutions (with respect to $\nu$ or $\nu$ and $\alpha$,
respectively) coincide in the sense that $ $the optimal value $\alpha^{*}\to0$
and the optimal value of $\nu$ is the same. Moreover ${\cal E}_{PS'}\ge{\cal E}_{PS}$.
When the filling is increased to, say, $n=0.4$ the situation is similar
with the exception of a small region ($u\lesssim0.8$ for $z=0$ or
$u\lesssim2.8$ for $z=0.9$) that can be considered to be outside
the scope of the strong-coupling approach. At $n=0.6$, the PS' solution
can be ignored for $u\gtrsim3.2$ (a result checked at $z=0$ and
$z=0.1$) or $u\gtrsim5.2$ (as checked at $z=0.5$ and $z=0.9)$.
Thus, close to half-filling the PS' becomes relevant also at intermediate
values of $u$ and we have examined it in more detail.

Let us fix $n=0.9$; in the symmetric case $z=0$ the optimal value
of $\alpha$ remains at its maximum $\alpha^{*}=n/2$ for $u<32$
where it starts to decrease to reach $\alpha^{*}=0$ at $u\cong44.3$;
the PS' state has a lower energy density with respect to the PS one
for $u\lesssim45.8$. For positive $z$ as long as $z\le0.43$ the
PS' solution is never better than the ones considered before. When
$z$ increases further the PS' state is favored over the PS or even
the SF one; the region at large $z$ and moderate $u$ where $\vert\Psi\rangle_{PS'}$
has a lower energy is characterized by the fact that $\alpha^{*}=n/2$
and $\nu^{*}=n/2$ so that $p=\ell_{\downarrow}=1$ meaning that all
the down spins are isolated from each other. This configuration resembles
the trimer crystal phase found in ref. \cite{KCR2009}, with a mixture
of hardcore bosons with attractive interaction and fillings 1/3 and
2/3, which is equivalent to a repulsive case with balanced species
and total filling 2/3 when a particle-hole transformation is performed.

\section{Conclusions\label{sec:conc}}

Our study, which combines analytical calculations in the strong- and
weak-coupling regimes and DMRG simulations both for attractive and
repulsive interaction, sheds some light on three qualitative and quantitative
questions that are still open in the literature of the 1D AHM:

\begin{enumerate}
\item The shape of the transition line from SS to CDW dominant correlations
for $U<0$ is filling-dependent and re-entrant in some cases (see
Fig. \ref{fig:sscdw}); 
\item Phase separation and phase segregation take place close to the Falicov-Kimball
limit above an interaction value $U_{c}$ which depends on the population
in such a way that it approaches zero in the small density regime.
Furthermore, transitions between phase separation and phase segregation
at varying interaction take place at a nearly constant asymmetry; 
\item For small asymmetry, close to the Hubbard limit $t_{\downarrow}\lesssim t_{\uparrow}$,
the SF-PS transition takes place at larger and larger values of $U$;
Eq. (\ref{eq:PSlargeu}), obtained in the framework of a variational
strong-coupling argument, indicates that an arbitrarily small asymmetry
is sufficient, at very large repulsions, to create a phase separated
state which destroys the spinless fermion-like ground state of the
Hubbard model. 
\end{enumerate}
\begin{acknowledgments}
We are grateful to Giuseppe Morandi, Arianna Montorsi and Alberto
Anfossi for useful and interesting discussions. This work is partially
supported by Italian MIUR, through the PRIN grant n. 2007JHLPEZ. 
\end{acknowledgments}

\section*{Appendix: Free spinless fermions with open boundary conditions}

The eigenfunctions of the hopping operator $-t\sum_{j=1}^{L-2}(c_{j}^{\dagger}c_{j+1}+{\rm h.c.})$
have the form\[
\varphi_{m}(j)=\sqrt{\frac{2}{\ell+1}}\sin(k_{m}j)\;,\;\; k_{m}=\frac{\pi m}{\ell+1}\;,\;\; m=1,\dots,\ell\]
 and the dispersion relation is formally the same as in the case of
PBC $\epsilon(k_{m})=-2t\cos(k_{m})$ so\[
\kappa=-2t\sum_{m=1}^{p}\cos\left(\frac{\pi m}{\ell+1}\right)=t\left\{ 1-\frac{\sin\left[\frac{\pi(2p+1)}{2(\ell+1)}\right]}{\sin\left[\frac{\pi}{2(\ell+1)}\right]}\right\} \]
 where $p\le\ell$ is the number of particles.

To compute the average density on the $j$-th site $\langle n_{j}\rangle$
we pass to the creation/annihilation operators in $k$-space\[
c_{j}=\sqrt{\frac{2}{L+1}}\sum_{m=1}^{\ell}\sin\left(\frac{\pi m}{\ell+1}j\right)\tilde{c}_{k_{m}}\]
 \[
\langle n_{j}\rangle=\frac{2}{L+1}\sum_{m,m'}\sin\left(\frac{\pi m}{\ell+1}j\right)\sin\left(\frac{\pi m'}{\ell+1}j\right)\times\]
 \[
\times\langle0\vert\tilde{c}_{k_{1}}\dots\tilde{c}_{k_{p}}(\tilde{c}_{k_{m}}^{\dagger}\tilde{c}_{k_{m'}})\tilde{c}_{k_{p}}^{\dagger}\dots\tilde{c}_{k_{1}}^{\dagger}\vert0\rangle\]
 The only non-vanishing possibility within the matrix element for
the vacuum $\vert0\rangle$ filled up to the momentum $k_{p}=\pi p/(\ell+1)$
is $m'=m$, so we have the characteristic function of the Fermi sea
$n_{k_{m}}$\[
\langle n_{j}\rangle=\frac{2}{\ell+1}\sum_{m=1}^{p}\sin^{2}\left(\frac{\pi m}{\ell+1}j\right)\]
 \[
=\frac{p}{\ell+1}-\frac{1}{\ell+1}\sum_{m=1}^{p}\cos\left(\frac{2\pi m}{\ell+1}j\right)\]
 \begin{equation}
=\frac{2p+1}{2(\ell+1)}-\frac{\sin\left[\frac{\pi(2p+1)}{\ell+1}j\right]}{2(\ell+1)\sin\left(\frac{\pi}{\ell+1}j\right)}\label{eq:dSF}\end{equation}
 The density at the edge is obtained by setting $j=1$.

\end{document}